\documentstyle[a4, epsf, psfig, doublespace]{article}
\begin{document}
\title{Density waves of granular flow in a pipe using lattice gas
automata}
\author{Gongwen Peng \thanks{On leave from Institute of Physics,
Academia Sinica, Beijing, China.} and
Hans J. Herrmann \\ HLRZ, KFA J\"ulich, D--52425
 J\"ulich, Germany}
\date{\today}
\maketitle
\begin{abstract}
\noindent
We use a lattice gas automaton modelling the formation of density
waves of granular flow through a vertical pipe. It is found
that both the dissipation and the roughness of  walls of the pipe
are essential to the emergence of density waves.
The density waves can only be observed when the average
density of the system is in a certain range.
The power spectra of density fluctuations in one  region in the pipe
follow,  apart from a sharp peak corresponding to the density
wave,  a power
law spectrum $1/f^\alpha$ with $\alpha$ close to $4/3$.\\
\end{abstract}

\noindent
\raisebox{17.5cm}[][]{preprint HLRZ 72/93}
\newpage
\noindent
Granular materials exibit many unusual phenomena such as size
segregation \cite{Williams, Haff, Rosato, Devillard}, heap formation and
convection cells under vibration \cite{Farady, Evesque, Taguchi,
Gallas},  and anomalous sound propagation
\cite{Liu, Jaeger}. Even in simple geometries like hoppers and pipes,
their flow under gravity still shows complex dynamics \cite{Baxter,
Poschel}. Experiments \cite{Baxter, Poschel} and molecular dynamics
(MD) simulations \cite{Poschel, Ristow, Jysoo} show that the granular
particles do not flow uniformly but rather form density waves (or
shock waves) where regions of high density travel with velocity
different from that of the average velocity. In the experiment of flow
in a hopper, Baxter et. al. \cite{Baxter} found that density waves
only exist when rough sands were used. To understand the density waves
in granular flow, attempts have been made by computer simulations with
MD \cite{Poschel, Ristow, Jysoo} and kinetic wave approach \cite{Jysoo,
Leibig}. The mechanism for the density waves is, however, not really
clarified so far.\\
The present work is to study this problem from another point of view,
namely lattice gas automaton (LGA) which was first introduced as a
novel alternative to traditional methods for numerically solving the
Navier--Stokes equation \cite{FHP}. As a sort of primitive molecular
dynamics system it offers the advantage of guaranteed numerical
stability coupled with extreme computational simplicity.
We are interested in the density waves in a vertical narrow pipe which
were  observed experimentally by P\"oschel \cite{Poschel} and
simulated with MD by him and later by Lee \cite{ Jysoo}. We
consider a
LGA at integer time steps $t=0,1,2,\cdots$ with $N$ particles located  at
the sites of a 2D triangular lattice which is $L$
sites long vertically and $W$ sites wide horizontally. Periodic
boundary conditions are used in the vertical direction while fixed
boundary conditions are set for the walls. At each site
there are seven Boolean states which refer to the velocities,
$\vec{v}_i (i=0,1,2,\cdots 6)$. Here $\vec{v}_i (i=1,2,\cdots 6)$ are
the nearest neighboring (NN) lattice vectors and
$\vec{v}_0=\vec{0}$  refers
to the rest (unmoving) state. Each state can be either empty or
 occupied by a
single particle. Therefore, the number of particles per site has a
maximal value of $7$ and a minimal value of 0. The time evolution of
LGA consists of a collision step and a propagation step. In the
collision step particles change their velocities due to collisions and
in the subsequent propagation step particles move in the directions of
their velocities  to the NN sites where they collide again.\\
The system is updated in parallel. Only the following specified
collisions can deviate  the trajectories of particles. All collisions
conserve mass and momentum.\\
Let us number the six bonds connected to a site counterclockwise, with
an index $i$, defined as  the integers (mod $6$), $i=1,2,\cdots 6$,
and label the rest particle with index $0$.  We consider only two-- and
three--body collisions.\\
For two--body collisions, we have:\\
(1) $(i, i+3)$ goes to $(i+1,i-2)$ and $(i-1,i+2)$ with equal
probability of $\frac{1}{2}$. Here $(i, i+3)$ means two particles
with opposite velocities  making
a head--on collision (this notation was also used in  \cite{FHP}). \\
(2) $(i,i+2)$ goes to $(0,i-2)$ with probability of $p$ and $(i+3,i-1)$
with probability of $1-p$. If $p$ is non--zero, this means that the
energy is dissipated due to  collision. This is the case for rough
granular particles.\\
For three--body collision, we have:\\
(3) $(i,i+2,i-2)$ goes to $(0,i,i+3), (0,i-1,i+2), (0,i+1,i-2)$ with
probability of $\frac{p}{3}$ for each, and $(i+3,i+1,i-1)$ with
probability of $1-p$.\\
The collision rules for moving particles with a rest particle involve
typical
 mechanisms of  granular flow \cite{Savage, Goldhirsch}.
Intuitively one can  understand them as follows. Rest particles in a
 region
will decrease the local granular tempurature which can be regarded
to be  the (kinetic) energy, causing a decrease in
pressure in that region. The resulting pressure gradient will lead to
a migration of particles into that region, increasing its density and
decreasing its pressure and granular temperature even more. That
means that  rest
particles will induce having more rest particles nearby. However, due
to  the
restriction of LGA that the rest state at one site  can at most be
 occupied
by one particle, we cannot simulate the above--mentioned mechanisms
 easily. For example, two moving particles colliding with a rest
particle
from opposite directions  can stop each other in accordance with
momentum conservation. But on each site only one rest particle is allowed.
Therefore, the collision should be taken by an off--site collision, i.
e., the two particles stay at rest on the NN sites where they
originally came from.
However, on these sites there may already exist other rest particles.
To
make things easy, we will still use the on--site collision but temporarily
allow more than one particle on a site during the collision.
Immediately after the collision, the extra rest particles
 hop to  NN sites randomly
until they find a suitable site with  no rest particle already sitting
there. Only in
this way can we incorporate the mechanisms mentioned above. The
collision rule with rest particles are as follows:\\
(4) $(i,0,i+3)$ goes to $(0,0,0)$ with certainty.\\
(5) $(i,0,i+2)$ goes to $(0,0,i-2)$ with certainty.\\
So far, we have not considered the gravity which is the driving force
of the flow. We simply incorporate it by the following rule:\\
(6) A rest particle decides to have a velocity along the direction of
gravity with probability  $g$, if the resulting state is empty at
that time. A moving particle colliding with a rest particle can change its
velocity by a unit vector along gravity with probability  $g$, if
the resulting state is possible on the trianglar lattice used.\\
The sites at  the walls of the system only have two directions
into which the  particles can move. So, the collision rule with the walls
reads:\\
(7) A particle colliding with the wall from one direction can be
bounced back with probability  $b$ and specularly reflected into
the other direction with probability  $1-b$. If $b=0$, the walls are
 smooth (perfect no--slip condition). Otherwise, the walls  have some
roughness.\\
We evolve the system according to the collision rules defined above.
The initial configuration of the system is set to be random in the
sense that each state (except the rest state) of each site is randomly
occupied according to a  preassigned average density $\rho$. In the
following we report the results made on systems with length $L=2200$
and width $W=11$. The lattice spacing is taken to be unit and the
triangular lattice has an axis parallel to  gravity and to the
walls.\\
Fig. 1(a) shows the time evolution of the density in the pipe measured
 every 80 time steps from $t=1$ to $t=40,000$ for $p=0.1, g=0.5,
b=0.5$. The average density of the system is $\rho=1.0$ (note that the
range of $\rho$ is between $0$ and $7$). The density plots are made as
follows. We divide the pipe along the vertical direction into $220$
bins  with equal length
of $10$ (total length $L=2200$) and count the number of particles
$n_i$ in   the $i$th bin. The
grayscale of each bin is a linear function of $n_i$.  The
$ n_i (i=1,2,\cdots)$ at a given time are plotted from left to right
while densities at different time steps are plotted from top to bottom
as time increases. Gravity is from left to right. We see that
initially the density is rather uniform and gradually regions of high
density
 are being formed out of the homogeneous system. A high density
region may also die out and two high density regions may merge to form
a single one.  It seems clear
that these are the same density waves (or shock waves) which were also
observed in
experiments \cite{Baxter, Poschel} and MD  simulations
\cite{Poschel, Ristow, Jysoo}. We also found that the width of the density
wave initially increases with time and then saturates after many  time
steps.
For most of the time, these density waves just travel with almost
constant velocity which  depends  on  the parameters
($p, g, b, \rho$) used. This constant velocity was also noted in MD
simulations \cite{Jysoo}.\\
The average density $\rho$ plays an important role in the formation of
density waves. We found that the density waves can only be observed in
a certain range of $\rho$. This range is almost independent of
$p, g$ and $b$ and is approximately between  $0.6$ and $ 1.6$.\\
The difference between the "granular gas" and a regular gas is in the
inherent dissipative nature of the elementary collision processes. Here
we include the dissipation through the parameter $p$. If we switch off
$p$,
no dissipation is present. In such a case, strikingly, the density
waves do not form. This is shown in fig. 1(b) which is
otherwise the same as fig. 1 (a) except for $p=0$. From this we can
conclude that the dissipation among the particles is essential to the
formation of density waves (this was experimentally observed by Baxter
et al
in a hopper \cite{Baxter}).\\
One  advantage of simulations is that one can easily control and
modify
the dynamical process. From $t=1$ to $t=40,000$ we switch off
the dissipation, and  the results are in fig.1(b). Now starting from
$t=40,000$, we switch  the dissipation on by setting $p=0.1$, and we
obtain  the
results of fig. 1 (c) where density waves are observed again. This
reveals that even for a minute   degree of
 dissipation  (provided  $p$ is non--zero), its mere
existence gives rise to  significantly different physics as compared
to that of regular gases and liquids.\\
The roughness of walls is also essential to the density wave
formation. When we turn off the roughness parameter $b$ ($b=0$), we
observe no density waves either. This is shown in fig. 1 (d).\\
To characterize the density fluctuations in a certain region with time,
we calculate their power spectra. We did this in a system of length
$L=220$ and width $W=11$. We recorded  the number of particles in a
vertical region of length $10$ every $10$ time steps. The dynamical
process is performed for very long time steps so that we obtain $256K
(1K=1024)$ data to analyse for each power spectrum. We first substract
the mean value from the data, otherwise there would be  a huge peak at
$f=0$ in
the power spectra. We calculated the
spectra using a standard  FFT routine. To get better statistics, average
process has been used. We broke the time series of
$256K$ points into $S$ segments of $M$ points each. On each segment an
FFT  was
performed  using a Parzen window \cite{book} and the
powers of the resulting spectra were averaged. Here we used $S=4$. A
representative power spectrum is shown in fig. 2 for $p=0.5, g=0.5,
b=0.5, \rho=1.0$. The frequency is in an arbitrary unit and  can be
related to the real time period (we check this relation by using the
same  program to
analyse a series of
exactly periodic data with the same number of points). The frequency
$f=1000$ in fig.2 corresponds
to a time period of    $T=328$. In fig. 2 we observe that a sharp peak exists
around $f=596$. The time period of this peak is
$T=\frac{328000}{596}=550$ and  corresponds to a wave  velocity
of $v=\frac{L}{T}=\frac{220}{550}=0.4$. This value coincides very well
with the velocity we measured for the density waves directly from the
time--evolution plots
of density (similar to fig.1(a) but for $p=0.5$).  That is to
 say, the
higest density region travelling periodiclly in the system (see
fig.1(a)) contributes to the power spectrum a sharp peak.   From fig.2
one sees that apart from this peak there is  a power law regime where
the spectrum falls off as $1/f^\alpha$. The  line in the log--log
plot of fig.2 represents a least square fit to the points between
$f=40$ and $f=1000$ after substracting the sharp peak. The exponent is
$\alpha=1.33\pm 0.02 (\simeq 4/3)$. Our ability to study the behavior in
the very low frequency regime ($f<40$) was limited by the long time
series of data required. The exponent $\alpha$ is found to be independent
on the parameters used but the position of the sharp peak (therefore
the velocity of the density wave) does depend on the parameters.\\
The picture revealed by fig. 2 is rather clear. In fig.1(a) we
observed the high density region travelling periodically due to the
periodic boundary condition.
This wave is very strong and most clearly distinguished from the rest.
This is why the peak  in fig.2  is so high. Apart from this wave, there
are also other waves with a broad range of frequencies (or travelling
velocities). The spectrum  of these waves follows a power law distribution
$1/f^\alpha$ with $\alpha$ close to $4/3$, which might be associated
to the dissipation instability discussed in \cite{Savage, Goldhirsch}.\\
In conclusion, by using a lattice gas model we observed the density
waves found in experiments and MD simulations. The density waves  only
exist in a range of average
densities.  Both the
dissipation among the granular particles and the roughness of the walls
of pipes are essential to the formation of such travelling waves which
seem to be simialr to the kinetic waves also observed in traffic jams
\cite{Jams}. The
density fluctuations follow, apart from  a sharp peak corresponding to the
density wave discussed above,  a $1/f^\alpha$ power spectra with
$\alpha $ close to $4/3$. Power--law spectra have also been observed
in experiments \cite{Baxter, Poschel} and MD simulations \cite{Ristow}
and we conject that they are a direct consequency of the dissipation
instability \cite{Savage, Goldhirsch}. \\
We acknowledge  the members of HLRZ Many Body Group for stimulating
discussions. Special thanks to Jan Hemmingsson for
constant conversation.\\

\newpage
\noindent
{\bf Figure Captions:}\\
\noindent
Fig.1: Time evolution of the density $n_i \{i=1,2,\cdots 220\}$
in the $220$ bins in the pipe of $L=2200$, $W=11$ and $\rho=1.0$.
 Densities at a given time are
plotted from left to right (direction of gravity) while
densities  at different time steps are plotted from top to bottom
 (direction of time increase)
every 80 time steps during
$40,000$ time steps.
The  grayscale of each bin is a linear function of $n_i$. Darker
regions  correspond to higher densities. (a) [topleft] $p=0.1, g=0.5, b=0.5$;
(b) [topright]  $p=0, g=0.5, b=0.5$;
(c) [bottomleft] continued from (b) but with dissipation switched on
by setting
$p=0.1$;  (d) [bottomright] $p=0.1, g=0.5, b=0$.\\
\noindent
Fig.2: Power spectrum of the time series of the density fluctuation
inside
a certain region in a pipe of $L=220$ and $W=11$. Parameters used here
are $p=g=b=0.5, \rho=1.0$. The straight line is  least square
fit with  a slope of $-1.33\pm0.02$.\\
\begin{figure}[p]
\centerline{
\epsfxsize=6.0cm
\epsfbox{fig_1_a.ps}
\hfil
\epsfxsize=6.0cm
\epsfbox{fig_1_b.ps}}
\centerline{
\epsfxsize=6.0cm
\epsfbox{fig_1_c.ps}
\hfil
\epsfxsize=6.0cm
\epsfbox{fig_1_d.ps}}
\caption {}
\end{figure}

\begin{figure}[p]
\centerline{
\epsfxsize=10.0cm
\epsfbox{fig_2.ps}}
\caption {}
\end{figure}

\end{document}